\documentclass[pdftex,twocolumn,epjc3]{svjour3}         

\RequirePackage[T1]{fontenc}

\smartqed  

\RequirePackage{graphicx}
\RequirePackage{mathptmx}      
\RequirePackage[numbers,sort&compress]{natbib}
\RequirePackage[colorlinks,citecolor=blue,urlcolor=blue,linkcolor=blue]{hyperref}

\usepackage{graphicx}
\graphicspath{{fig/}}
\usepackage{siunitx}
\usepackage{hyperref}
\usepackage{epsfig}
\usepackage{booktabs}
\usepackage{multirow}
\usepackage{slashbox} 
\usepackage{lipsum}
\usepackage{mathtools}
\usepackage{cuted}
\hypersetup{hidelinks,colorlinks=false,breaklinks=true,urlcolor=ocre}

\journalname{Eur. Phys. J. C}

\begin{document}

\title{Development of Planar P-type Point Contact Germanium Detectors for Low-Mass Dark Matter Searches}


\author{W.-Z. Wei\thanksref{addr1}
        \and
        H. Mei\thanksref{addr1} 
        K. Kooi\thanksref{addr1} 
        D.-M. Mei\thanksref{e1,addr1}, 
        J. Liu\thanksref{addr1},
        J.-C. Li\thanksref{addr1},
        R. Panth\thanksref{addr1} and
        G.-J. Wang\thanksref{addr1}
        }

\thankstext{e1}{Corresponding Author: Dongming.Mei@usd.edu}

\institute{Department of Physics, The University of South Dakota, Vermillion, South Dakota 57069\label{addr1}}

\date{Received: date / Accepted: date}
\maketitle

\begin{abstract}
 The detection of low-energy deposition in the range of sub-eV through ionization using germanium (Ge) with a bandgap of $\sim$0.7 eV requires internal amplification of the charge signal. This can be achieved through high electric field which accelerates charge carriers which then interact and generate more charge carriers. The minimum electric field required to generate internal charge amplification is derived for different temperatures. A point contact Ge detector provides extremely high electric field in proximity to the point contact. We show the development of a planar point contact detector and its performance. The field distribution is calculated for this planar point contact detector. We predict the required electric field can be achieved with a point contact detector.  

\end{abstract}

\section{Introduction}
\label{sec:intro}
 From an experimental point of view, the detection of low-energy deposition from low-mass dark matter at MeV-scale remains a daunting challenge. Until now, the devices that can meet the energy threshold of $<$ 10 eV required for detecting light dark matter in the MeV-scale range are still in their infancy~\cite{mei, supercdms1, supercdms2, edelweiss, cresst, damic, sensei1, sensei2, collar}.  
 
 
 High-purity Ge (HPGe) detectors are well known for providing excellent energy resolutions over a wide energy range, and thus have been widely used in nuclear and particle physics. A low capacitance Ge detector can be realized with a p-type point contact (PPC) geometry, which allows the detector to achieve the capacitance on the order of 1 pF and can be operated with sub-keV energy threshold~\cite{barb}. Thus, the PPC Ge detector technology is able to offer both excellent energy resolution and relatively low energy threshold for ultra-low background experiments such as GeV-scale dark matter and axion interactions. This technology has been exploited by two leading $^{76}$Ge-based neutrinoless double beta decay experiments, the M{\sc{ajorana}} D{\sc{emonstrator}}~\cite{mjd1} and GERDA~\cite{gerda}. However, some physical processes requiring an extreme low-energy threshold of $<$ 10 eV such as the detection of MeV-scale dark matter is still unreachable with existing PPC Ge detectors whose energy threshold are hundreds of eV due to the limitation from the read-out electronic noise. Most recently, by the introduction of ultra-low vibration mechanical cooling and wire bonding of a complimentary medal-oxide semiconductor (CMOS) charge sensitive preamplifier to a sub-pF PPC Ge detector, Barton et al.~\cite{barton} have demonstrated electronic noise of 39 eV-FWHM (full width at half maximum) at 43 K. Thus, PPC Ge detectors have great potential to achieve an extreme low energy threshold if the detectors can be operated at a even lower temperature such as liquid helium temperature ($\sim$4 K)~\cite{wei2020}. 
 
\begin{figure} [htbp]
  \centering
  \includegraphics[width=0.4\linewidth]{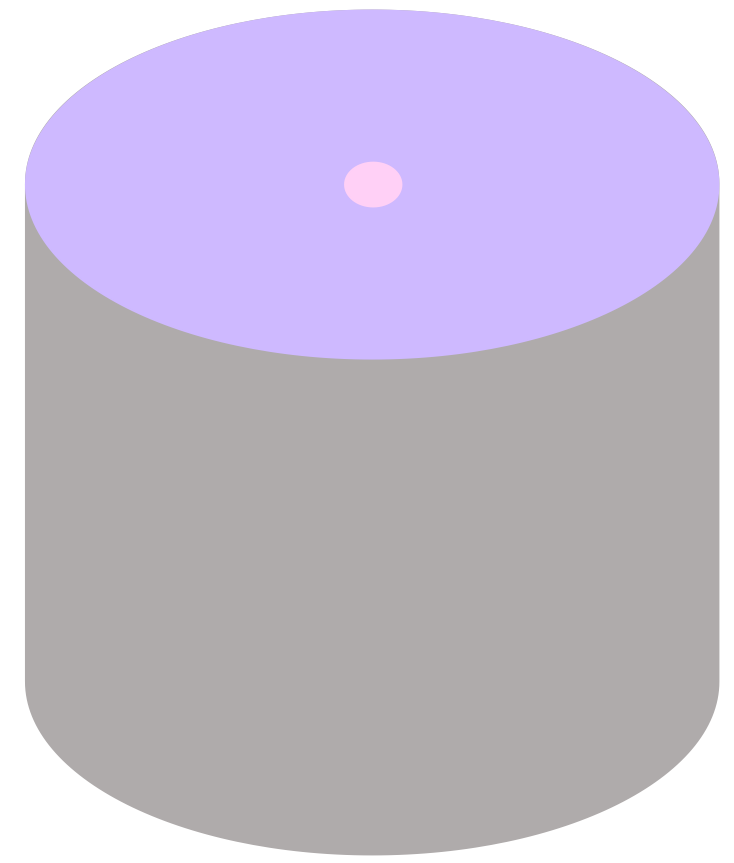}
  \includegraphics[width=0.4\linewidth]{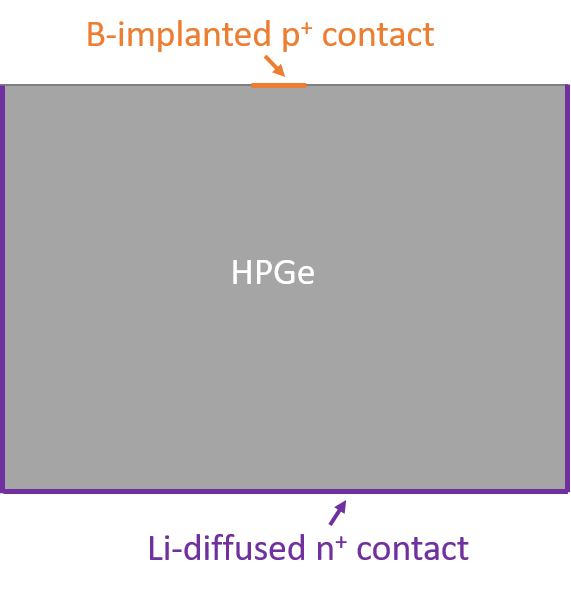}
  \caption{Cylindrical PPC Ge detector. Left: Sketch of the full detector~\cite{mjd2}; Right: Cross section. The p$^{+}$ contact is shown in orange and the n$^{+}$ contact is shown in purple.}
  \label{fig:coa-ppc}
\end{figure}

\begin{figure} [htbp]
  \centering
  \includegraphics[width=0.6\linewidth]{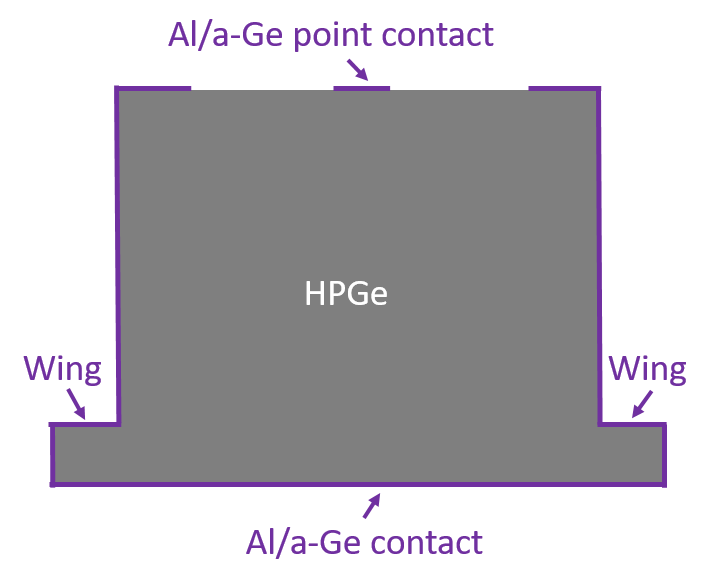}
  \caption{Cross-sectional drawing of a planar PPC Ge detector. The point and outer surface contacts are in purple and they are both made of a-Ge contact with Al on the top.}
  \label{fig:pla-ppc}
\end{figure}

 Traditionally, as shown in Fig.~\ref{fig:coa-ppc}, a PPC Ge detector is made of a cylinder of HPGe crystal in which B-implanted p$^{+}$ and Li-diffused n$^{+}$ electrical contacts are fabricated on the point contact and outer surface contact, respectively. Compared with the traditional PPC Ge detector shown in Fig.~\ref{fig:coa-ppc}, a planar PPC Ge detector with a-Ge contacts, as shown in Fig.~\ref{fig:pla-ppc}, can be fabricated much easier due to the two following reasons: (1) four wings in a planar geometry simplify the handling process during the detector fabrication; and (2) fabrication of a-Ge contacts is much simpler than the fabrication of traditional B-implanted p$^{+}$ and Li-diffused n$^{+}$ electrical contacts~\cite{amman,wei2020}. If a planar PPC detector shown in Fig.~\ref{fig:pla-ppc} is able to achieve a similar detector performance (low detector capacitance and excellent energy resolution) as a conventional PPC detector shown in Fig.~\ref{fig:coa-ppc}, due to its relative simpler fabrication process, it would be more convenient to use a planar PPC detector instead of a cylindrical PPC to explore some interesting physics processes such as charge trapping due to shallow level impurities and charge lost due to deep level impurities in Ge~\cite{tyl}. In addition, it could also be used as a drift detector for medical imaging~\cite{cas,kem2,gat}.

In order to investigate the possibility of observing internal charge amplification with a planar PPC Ge detector, we have fabricated such a detector in our lab at University of South Dakota (USD). The geometry of the detector is the same as shown in Fig.~\ref{fig:pla-ppc} and the detector was made with the Ge crystal grown by us at USD. The detector has been carefully characterized to examine its performance including leakage current, detector capacitance and energy resolution. Based on the detector performance, we have conducted a numeric calculation to study the electric field distribution inside the detector to find out if the required electric field for internal charge amplification can be achieved with this detector. In this paper, we first show the theoretical requirements for internal charge amplification in Sect.~\ref{sec:field}. The details about the fabrication and characterization of a custom-build planar PPC Ge detector are presented in Sect.~\ref{sec:fab_char}, followed by the discussion and the expectation of detector performance for observing internal charge amplification in Sect.~\ref{sec:dis}. Finally, the conclusions are summarized in Sect.~\ref{sec:conl}.

\section{Theoretical Requirements for Internal Charge Amplification}
\label{sec:field}
The initial charge carriers created by external energy deposition in Ge are through ionization and excitation of Ge atoms or impurity atoms in Ge. The ionization and excitation of Ge atoms to create the secondary charge carriers require a charge carrier to have a minimum kinetic energy of $\sim$0.73 eV (the band gap of Ge~\cite{mei}) at liquid nitrogen temperature (77 K). 
The main remaining impurities in the Ge after zone refining and crystal growth are boron, aluminum, gallium and phosphorus~\cite{mei}. They are all shallow impurities since their ionization energies are $\sim$0.01 eV~\cite{rwrw}. Charge carriers or phonons with energies larger than 0.01 eV can easily ionize or excite impurities from the donor level or acceptor level to create charge carriers. Therefore, the minimum energy required to generate a charge carrier and a charged impurity through ionization and excitation of impurities in Ge is $\sim$0.01 eV. 

Under an electric field, a charge carrier will be accelerated to gain a kinetic energy of $\Delta E=\frac{1}{2}m^{\ast}v_{d}^{2}$, where $m^{\ast}$ is the effective mass of the charge carrier and $v_{d}$ is the drift velocity. During the drifting process, the charge carrier will lose energy by interacting with the crystal lattice which manifests as emitted phonons~\cite{BNVT}.
These phonons are so-called Neganov-Luke phonons produced by charge transport~\cite{pnlu}. As a result, the emission of Neganov-Luke phonons will reduce the kinetic energy of the charge carrier. 
For a given drifting length, the energy conservation requires the work done on the charge carrier by the externally applied electric field to be equal to the sum of the kinetic energy gained by the charge carrier and the energy loss to the emission of phonons. This is to say that the work done on the charge carrier by the externally applied electric field is always balanced. This energy balance is performed through a competing process between the energy loss to the charge carriers scattering off the orbital electrons and the energy loss to the charge carriers scattering off the lattice~\cite{mei}. The former will ionize the orbital electrons to create additional charge carriers and the latter will cause the emission of phonons. To obtain internal charge amplification,  one prefers the energy loss to the charge carriers scattering off the orbital electrons. This requires the mean scattering length (the mean ionization length) for charge carrier scattering off the orbital electrons to be smaller than the mean scattering length for charge carriers scattering off the lattice.

\begin{figure} [htbp]
  \centering
  \includegraphics[angle=0,width=9.cm]{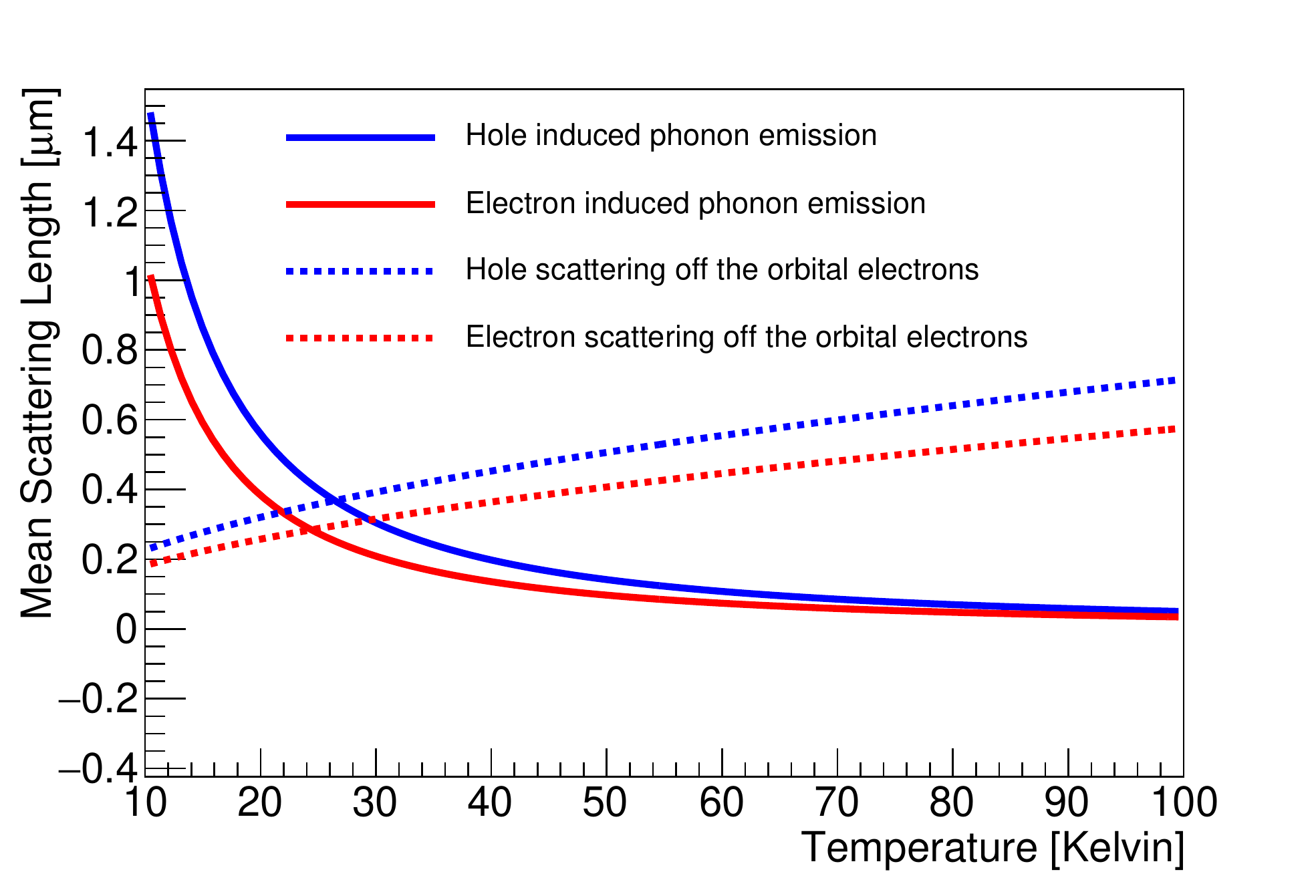}
  \caption{The mean scattering length versus temperature for charge carriers scattering off the orbital electrons and charge carriers scattering off the lattice, respectively at zero electric field. The solid blue line shows the mean scattering length of holes interacting with lattice for emitting phonons. The solid red line shows the mean scattering length of electrons interacting with lattice for emitting phonons. The dashed blue line shows the mean scattering length of holes interacting with orbital electrons. The dashed red lines shows the mean scattering length of electrons interacting with orbital electrons.  }
  \label{fig:msl}
\end{figure}

Under zero electric field, the ionization is a thermal process. Therefore, the mean scattering length depends only on temperature. For charge carriers scattering off the orbital electrons, the mean scattering length ($L_{th}$) can be calculated using $L_{th}$ = $v_{th}$$\times$$\tau_{th}$, where $v_{th}$ is thermal velocity of charge carriers and $\tau_{th}$ = $m^{\ast}\frac{\mu_{th}}{e}$ is the lifetime of charge carriers, with $\mu_{th}$ the charge mobility. For charge carriers scattering off the lattice, the mean scattering length ($\lambda_{th}$) can be obtained using $\lambda_{th}$ = $v_{s}\tau_{ph}$, where $v_{s}$ = 5.4$\times$10$^{5}$ cm/s is the speed of sound (phonon) in Ge and $\tau_{ph}$ = $m^{\ast}\mu_{ph}/e$, where $\mu_{ph}$ = $\frac{4.65\times10^{5}}{m^{\ast^{5/2}}}\cdot T^{-3/2}$ is the charge mobility due to acoustic deformation potential scattering~\cite{meihao} and $T$ is temperature. Figure~\ref{fig:msl} shows the mean scattering length as a function of temperature for ionization and phonon emission, respectively. 
It is clear that the mean scattering length for charge carriers scattering off the orbital electrons is smaller than that of charge carriers scattering off the lattice at temperature less than 30 K. To effectively generate more charge carriers with internal charge amplification, one would prefer to run the detector at a temperature less than 30 K. 

For drifting charge carriers under the electric field, the work done on a charge carrier can be expressed as:
\begin{equation}
\vec{F}\cdot\vec{L}=e\vec{E}\cdot\vec{L}=eEL,
  \label{6-9}
\end{equation}
where $F$ is the electric force, $L$ is the drifting path length along the direction of electric force, and $\vec{E}$ is the electric field. For one interaction length,  $L$ is equal to the mean scattering length for charge carriers scattering off the orbital electrons, that is  
\begin{equation}
L = v_{d}\cdot \tau=\mu E\frac{m^{\ast }\mu }{e}=\mu ^{2}E\frac{m^{\ast}}{e},
  \label{6-10}
\end{equation}
where $\tau$ is the relaxation time of a charge carrier and $\mu$ is the drift mobility. 

For temperature less than 30 K, the mean scattering length for charge carriers scattering off the orbital electrons is smaller than that for charge carriers scattering off the lattice, as shown in Fig.~\ref{fig:msl}. Therefore, the work done on a charge carrier by the applied electric field will mainly accelerate the charge carrier to gain sufficient kinetic energy and hence generate more e-h pairs through internal charge amplification. The minimum electric field required to trigger internal charge amplification can be roughly estimated using Eq.~\ref{6-9}. Since the minimum ionization energy of Ge atoms is $\sim$0.73 eV (the band gap energy of Ge) and the average ionization length is about 0.4 $\mu$m at zero electric field as shown in Fig.~\ref{fig:msl}, the minimum electric field required is just under 2$\times$10$^{4}$ V/cm for ionizing Ge atoms to initiate internal charge amplification. Similarly, the average ionization energy for shallow impurity atoms in Ge is about 0.01 eV. Thus, the minimum electric field required to trigger internal charge amplification through ionization of impurity atoms in Ge is roughly estimated to be less than 300 V/cm.  It is worth mentioning that if temperature is significantly lower than 30 K,  the average ionization length can be significantly larger than $\sim$0.4 $\mu$m under an applied electric field, as indicated in Fig.~\ref{fig:msl}. Therefore, the minimum electric field required to trigger internal charge amplification can be significantly lower than what is estimated for both ionization of Ge atoms and ionization of impurity atoms. However, the detailed discussion requires sophisticate models in calculating the average ionization length and is beyond the scope of this work.  

For temperature greater than 30 K, the average ionization length is larger than the mean scattering length for charge carriers scattering off the lattice. Hence, it is likely there will be energy loss to the emission of phonons before kicking off the internal charge amplification. Thus, energy conservation requires the work done by the electric field be equal to the sum of the gained kinetic energy and the energy of emitted phonons.  Therefore, the minimum electric field required to generate an e-h pair on Ge atoms or impurity atoms during a drifting process with a drifting length of $L$ can be calculated from the following relation:
\begin{equation}
eE\times L= \Delta E+\frac{L}{\lambda _{ph}}\times \bar{\Delta E _{ph}},
  \label{6-12}
\end{equation}
where $\lambda_{ph}$ = $v_{s}\tau$ is the mean scattering length for producing phonons, $\bar{\Delta E_{ph}}$ is the average energy of emitted phonons, which is estimated as $\bar{\Delta E_{ph}}$ = $\frac{hv_{s}}{a}$ = 0.04 eV, where $h$ is the Planck constant and $a$= 0.56 $nm$ is the lattice constant of Ge. 

Combining Eq.~\ref{6-10} and Eq.~\ref{6-12}, we have
\begin{equation}
eE\times\mu ^{2}E\frac{m^{\ast }}{e}=\Delta E+ \frac{\mu E}{v_s} \cdot \bar{\Delta E _{ph}}.
  \label{6-13}
\end{equation}
Rearranging the terms in Eq.~\ref{6-13}, we have 
\begin{equation}
m^{\ast }\mu ^{2}E^{2}-\frac{\mu}{v_s}\bar{\Delta E _{ph}}E-\Delta E=0.
  \label{6-14}
\end{equation}
For this quadratic equation, we have $E=\frac{-b\pm \sqrt{b^{2}-4ac}}{2a}$, where $a=m^{\ast}\mu^{2}$, $b=-\frac{\mu}{v_s}\bar{\Delta E _{ph}}$, and $c=-\Delta E$ is the energy required to generate an e-h pair in Ge. The minimum energy required to generate an e-h pair in Ge is $\sim$0.73 eV in Ge at liquid nitrogen temperature~\cite{wei2018}. 

Similarly, energy conservation during this process can be used to calculate the minimum electric field required to generate an e-h pair on impurity atoms in Ge using Eq.~\ref{6-13}. However, now $\Delta E$ is the ionization energy of shallow impurities in Ge ($\Delta E \geq$ 0.01 eV).

\begin{figure}
\centering
\includegraphics[angle=0,width=9.cm] {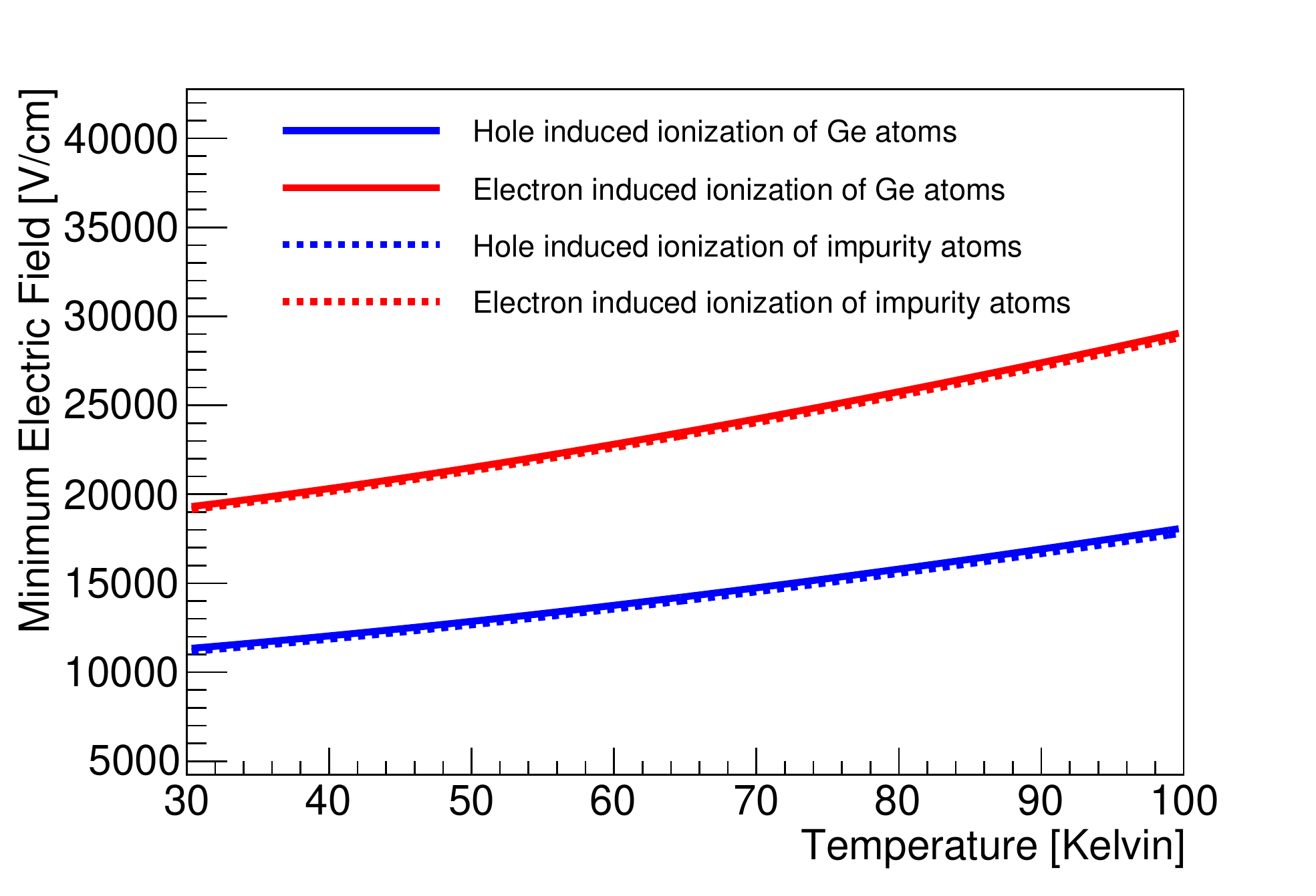}
\caption{ The minimum electric field required for internal charge amplification of Ge atoms and impurity atoms in a temperature range from 30 K to 100 K.}
\label{fig:6-1}
\end{figure}

When the temperature is above 30 K, the charge drift mobility in Ge is dominated by the contribution from phonon scattering mobility and neutral impurity scattering~\cite{meihao}. The Hall mobility for holes is often taken as $\mu_{h0}$ = 42,000 cm$^2$/(V$\cdot$s) and the Hall mobility for electrons is $\mu_{e0}$ = 36,000 cm$^{2}$/(V$\cdot$s) for Ge according to IEEE standard~\cite{ieee}. The relationship between the Hall mobility and the drift mobility is given by $\mu_{h}$ = $\mu_{h0}$/1.03 for holes and $\mu_{e}$ = $\mu_{e0}$/0.83 for electrons, respectively~\cite{ieee}. Utilizing Eq.~\ref{6-14} with $m^{\ast}$ = 0.21 m$_{0}$~\cite{mei} for holes and $m^{\ast}$ = 0.12 m$_{0}$~\cite{mei} for electrons, where m$_{0}$ = 511,000 eV/c$^2$ is the electron mass in vacuum, we obtain the minimum electric field as a function of temperature from 30 K to 100 K as shown in Fig.~\ref{fig:6-1}. 

It is clear that the minimum field required to generate an e-h pair in Ge is above 10$^4$ V/cm. There is a large difference in the minimum required field for generating more e-h pairs when drifting electrons and holes across the detector. A weak temperature dependence is also seen in Fig.~\ref{fig:6-1}. It shows that the minimum field required to generate more e-h pairs decreases as temperature decreases. This indicates that an internal charge amplification can be achieved at a lower electric field when the detector is operated at a lower temperature. At liquid nitrogen temperature (77 K), the required electric field for internal charge amplification must be obtained through a special geometry design at which high electric field is achievable in a designated detector region. We studied this hypothesis using a customized planar PPC detector.

\section{Operation and Characterization of A Custom-build Planar PPC Detector}
\label{sec:fab_char}

\subsection{Detector Fabrication Process}
\label{sec:fab}
\begin{figure} [htbp]
  \centering
  \includegraphics[width=0.439\linewidth]{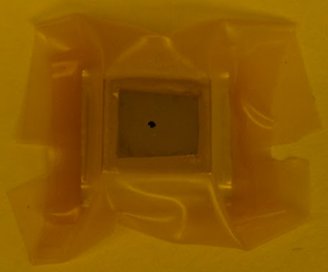}
  \includegraphics[width=0.425\linewidth]{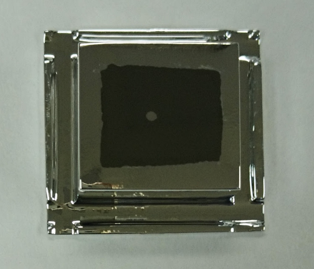}
  \caption{Left: The detector was covered with acid resistant tape and paint (picine) to form the outer contact and point contact, respectively. Right: The top view of the completed planar PPC Ge detector, USD-W06. The uneven edge of the outer contact was caused by cutting the acid resistant tape manually.}
  \label{fig:geo}
\end{figure}

To convert an HPGe crystal into a planar PPC detector, the fabrication process of depositing a-Ge and Al on all the detector surfaces is the same as in fabricating a planar detector from an HPGe crystal. Please refer to our previous work~\cite{wei2018,meng} for the fabrication details. After the electric contacts (a-Ge and Al) were deposited, as shown in Fig.~\ref{fig:geo} (left), an acid-resistant tape was used to cover the bottom, sides and a portion of the top surface to create the outer contact of the PPC detector. To form the point contact, an etch resistant paint (picine), as shown in Fig.~\ref{fig:geo} (left), was applied in the center of the top surface. The exposed Al on the point contact face was then removed using a diluted HF dip (100:1). The tape and paint were removed and the detector fabrication was completed. Figure~\ref{fig:geo} (right) shows the top view of the completed planar PPC detector, USD-W06, fabricated at USD. The diameter of the point contact is around 0.6 mm and the thickness of the detector is 9.4 mm. The width of the outer contact on the top surface of the crystal is around 6.0 mm. Other dimensions of this detector are presented in Table~\ref{tab:dim}.


\begin{table}[b]
  \centering
  \caption{The dimensions of the planar PPC Ge detector, USD-W06, fabricated at USD in this work.}
  \label{tab:dim}
  \begin{tabular}{p{2.5cm}p{2cm}p{2cm}}
    \hline
   &Length (cm) &Width (cm) \\ \hline
 Crystal Top & 1.37  & 1.37    \\ \hline
 Crystal Bottom& 2.12  & 2.08   \\ \hline
  \end{tabular}
\end{table}

\subsection{Detector Characterization}
\label{sec:test}
\begin{figure} [htbp]
  \centering
  \includegraphics[clip,width=0.6\linewidth]{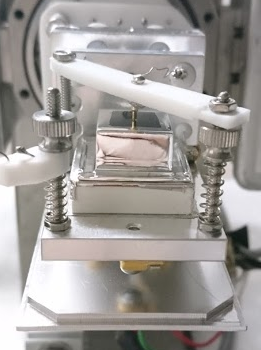}
  \caption{Shown is the detector USD-W06 loaded in a variable-temperature cryostat for characterization.}
  \label{fig:cryo}
\end{figure}

\begin{figure} [htbp]
  \centering
  \includegraphics[clip,width=8.5 cm]{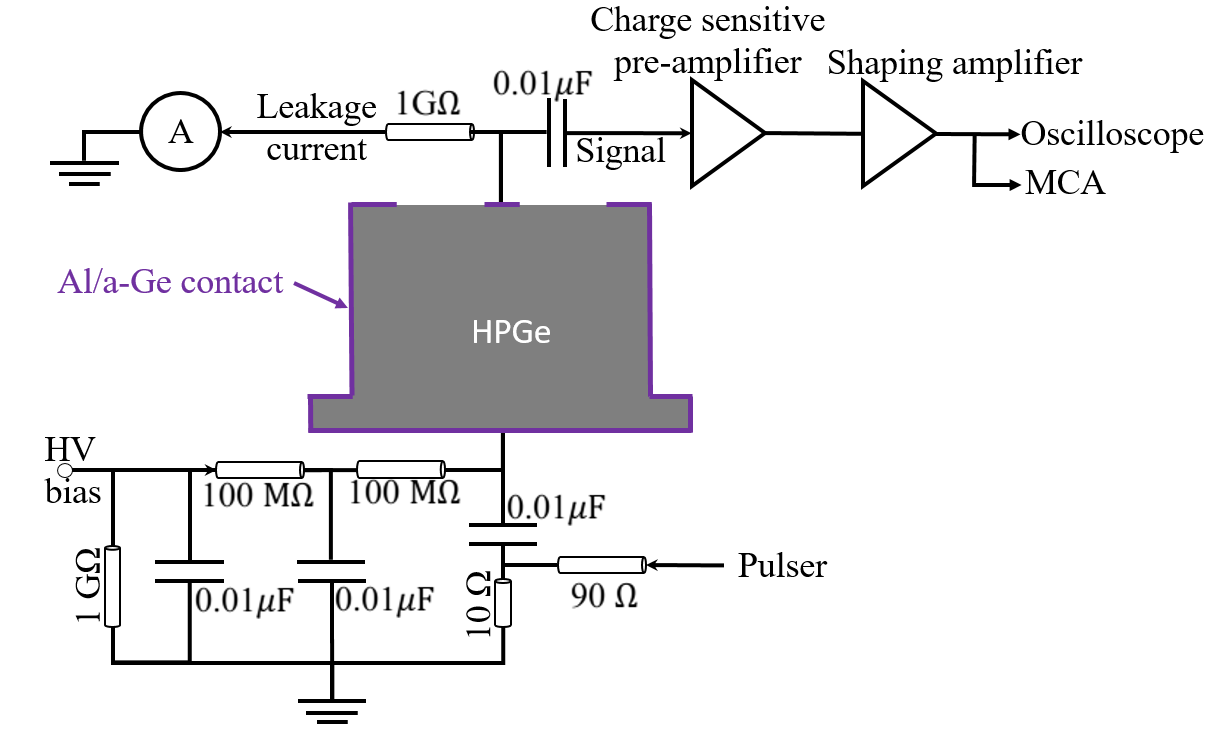}
  \caption{Electronic circuit for detector characterization.}
  \label{fig:electronic}
\end{figure}

After the detector was fabricated, it was loaded in a variable-temperature cryostat to conduct electrical and spectroscopy measurements, as shown in Fig.~\ref{fig:cryo}. This cryostat is provided by the Lawrence Berkeley National Laboratory (LBNL). Figure~\ref{fig:electronic} shows the electronics used for signal processing and conducting electrical and spectroscopy measurements in this work.

\subsubsection{Electrical Measurements}
The following two electrical measurements were conducted in this study: (1) the leakage current versus the applied voltage at 79 K; and (2) the detector capacitance as a function of the applied voltage at 79 K. For both electrical measurements, a positive bias voltage was applied to the bottom contact of the detector, USD-W06, and the signals were read out from the top point contact.

\begin{figure} [htbp]
  \centering
  \includegraphics[clip,width=9.cm]{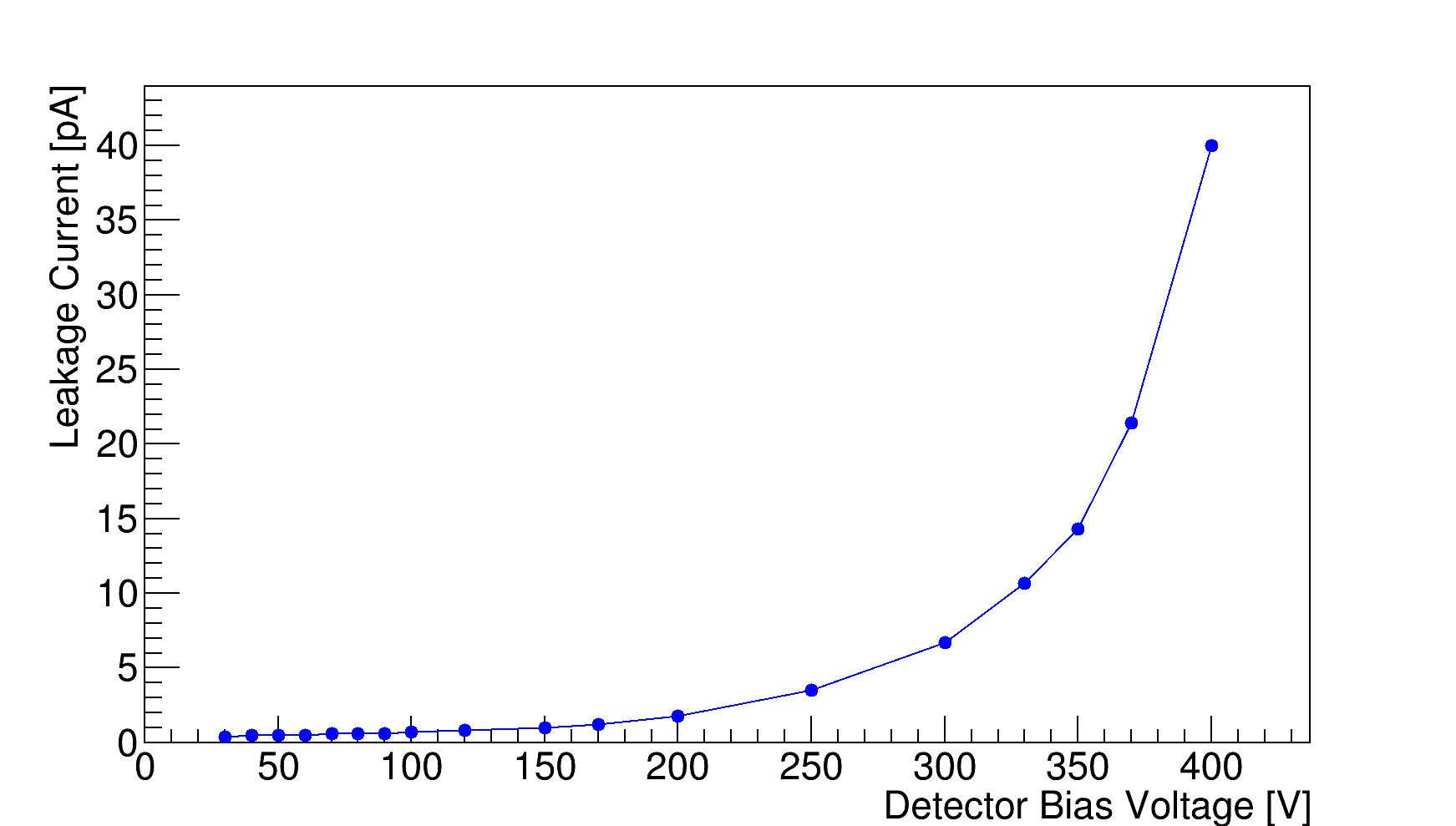}
  \caption{Measured leakage current as a function of bias voltage at 79 K for detector USD-W06. Low leakage current ($\sim$ 0.5 pA) at full depletion voltage (60 V) shows the good quality of a-Ge contacts.}
  \label{fig:IV}
\end{figure}

\begin{figure} [htbp]
  \centering
  \includegraphics[clip,width=9.cm]{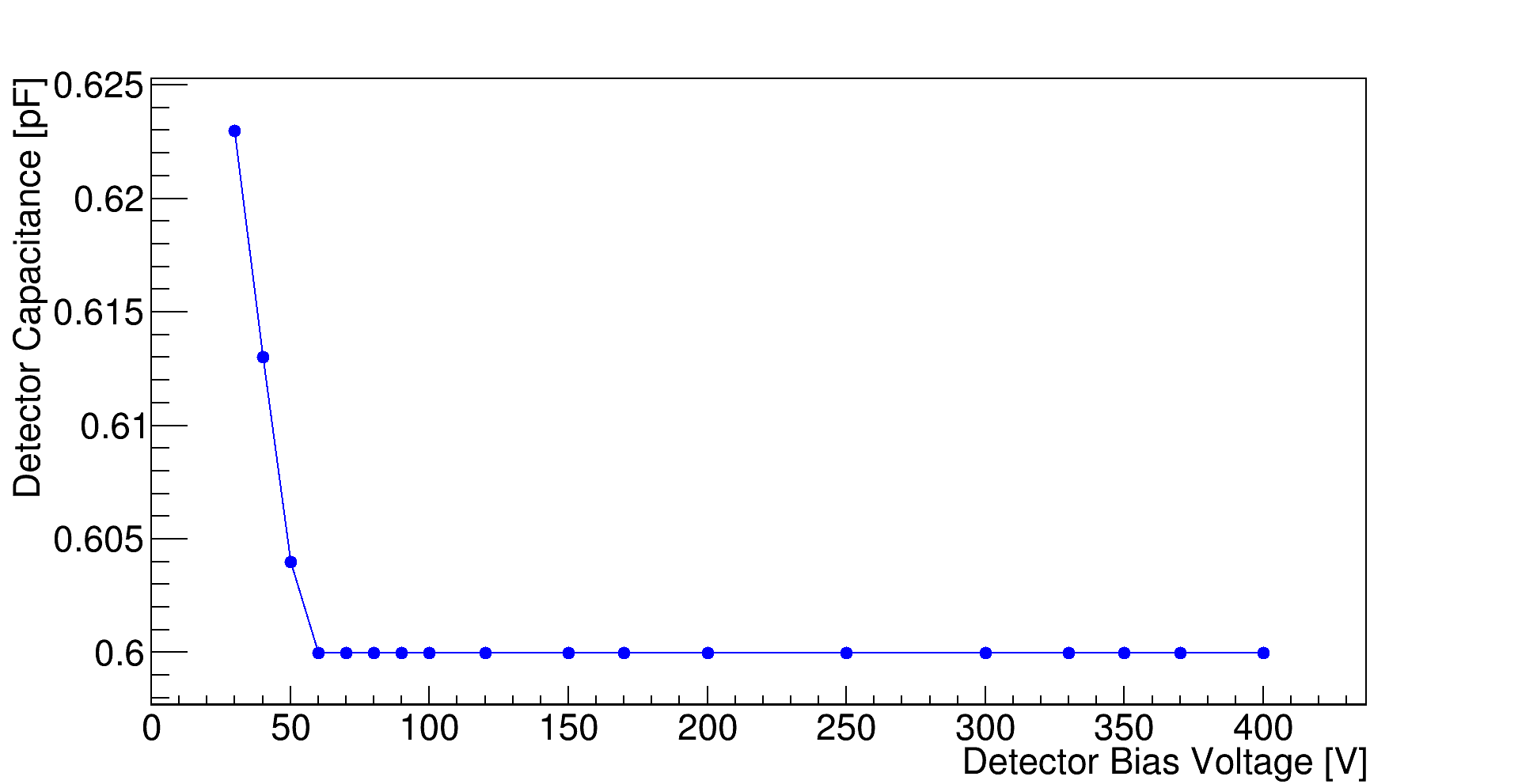}
  \caption{Measured detector capacitance as a function of bias voltage at 79 K for detector USD-W06. The full depletion voltage of the detector was determined to be 60 V since the detector capacitance became a constant after the detector was fully depleted.}
  \label{fig:CV}
\end{figure}

Figures~\ref{fig:IV} and ~\ref{fig:CV} present the measured leakage current and the detector capacitance as a function of the detector bias voltage at 79 K. As shown in Fig.~\ref{fig:CV}, detector USD-W06 is fully depleted at 60 V with leakage current at around 0.5 pA from Fig.~\ref{fig:IV}. Also, as shown in Fig.~\ref{fig:CV}, this planar PPC detector is able to reach a low capacitance of 0.6 pF. Since the signal to background is anti-proportional to the detector capacitance, a lower capacitance will enhance the signal to background ratio and hence has the potential to become a low-threshold detector if operated at an even lower temperature such as liquid helium temperature. 

\subsubsection{Spectroscopy Measurements}
\begin{figure} [htbp]
  \centering
  \includegraphics[clip,width=9.2 cm]{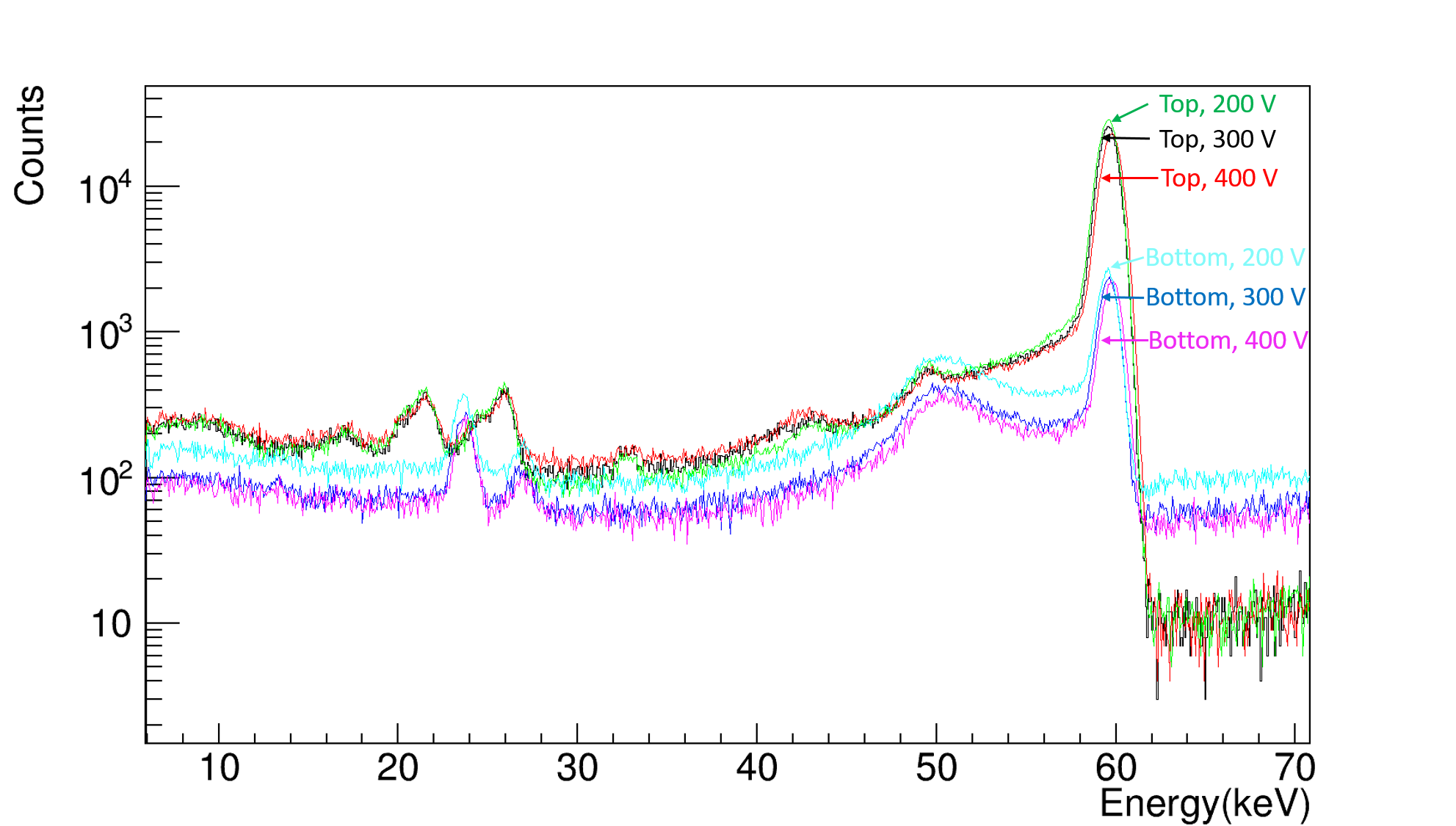}
  \caption{Energy spectra of a Am-241 source measured with the detector USD-W06. The spectra were taken at three voltages, 200 V, 300 V and 400 V with the source placed at the top and the bottom of the cryostat for each voltage.}
  \label{fig:spec}
\end{figure}

\begin{figure} [htbp]
  \centering
  \includegraphics[clip,width=9.2 cm]{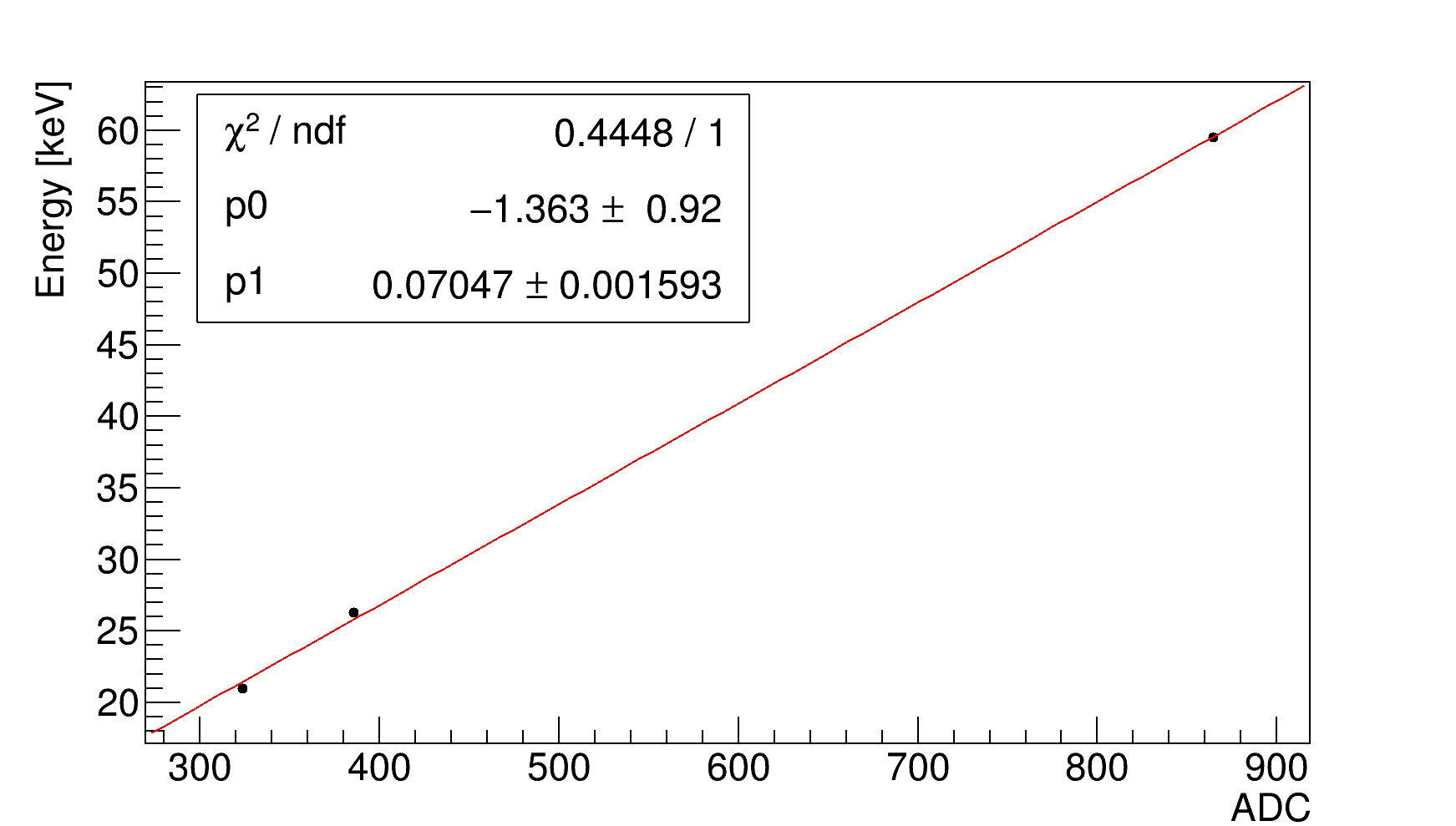}
  \caption{Converting the ADC channel number to the energy with linear calibration function of $E = p0 +p1\cdot A$, where $E$ is the energy, $A$ is the ADC channel number, and $p0$ and $p1$ are the fitting parameters.}
  \label{fig:cali}
\end{figure}

A radioactive source, Am-241, was used to conduct the spectroscopy measurements. In this study, there are two setups for the collection of the energy spectrum of Am-241: (1) the source was placed on the top of the cryostat, facing the top side of the detector; and (2) the source was placed at the bottom of the cryostat, facing the bottom side of the detector. For each setup, the energy spectra of Am-241 at 79 K were acquired with positive bias voltages of 200 V, 300 V and 400 V applied to the detector bottom with the detector configuration shown in Fig.~\ref{fig:cryo}. The signals were read out from the point contact using a charge sensitive pre-amplifier operated at room temperature followed by an ORTEC 671 shaping amplifier set to an optimized shaping time of 2 $\mu$s. Figure~\ref{fig:spec} shows the energy spectra of Am-241 taken at three voltages with the source placed at the two positions mentioned above. The data taking times for each of the applied voltages were 1.5 hours when the source was placed on the top of the cryostat, while the data taking times were 14.5 hours, 9 hours and 8 hours for the applied bias voltages of 200 V, 300 V and 400 V, respectively, when the source was placed at the bottom of the cryostat. The x-axis is the energy in keV, which is converted from the ADC count using three photon peaks from Am-241, $\sim$20.8 keV, $\sim$26.3 keV and 59.5 keV~\cite{am241}, with the applied voltage at 200 V and the source placed at the top of the cryostat. The calibration curve is shown in Fig.~\ref{fig:cali}. The linear fit function is: $E = p0 +p1\cdot A$, where $E$ is the energy in keV, A is the ADC channel number, and $p0$ and $p1$ are the fitting parameters shown in Fig.~\ref{fig:cali}. From Fig.~\ref{fig:spec} we can only see that the resolution in the region of $\sim$20 keV is worse than the 59.5 keV peak. From atomic data we know that for high-Z atoms, we expect clusters of X-rays. In addition, the energy region of $\sim$26.3 keV, a $\gamma$ ray from Am-241 source is contaminated by several X-rays from the $\sim$20 keV region, which are very close in energy. Therefore, the two peaks were not well resolved. 

\begin{figure} [htbp]
  \centering
  \includegraphics[clip,width=1\linewidth]{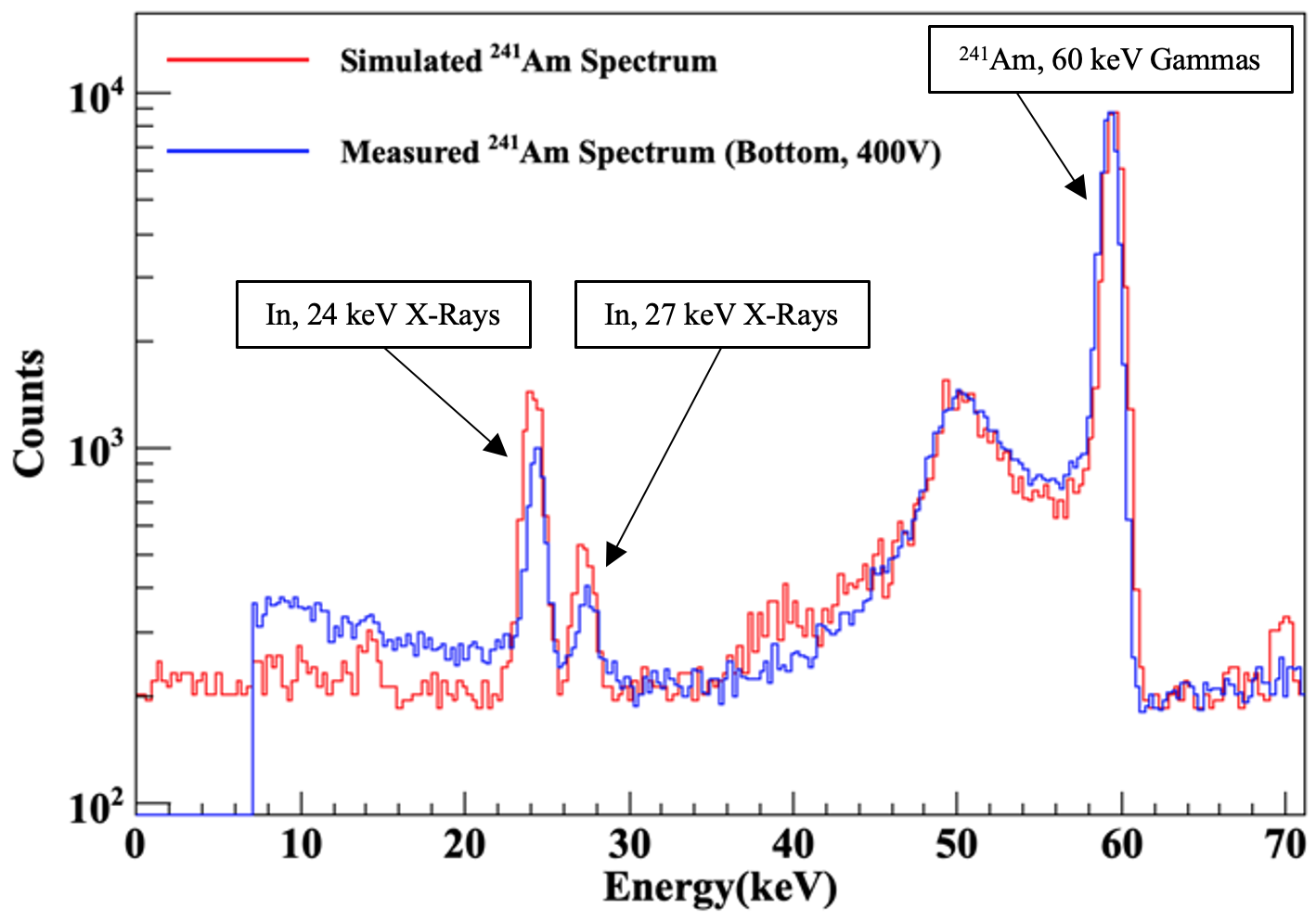}
  \caption{A comparison between the Monte Carlo simulation and the measurement when the source was placed at the bottom. The blue line is the measured energy spectrum and the red line is the simulated energy spectrum. The simulation confirms the fluorescent emission of X-rays (24 keV and 27.5 keV) from the inner material of indium. }
  \label{fig:geant4}
\end{figure}

It is worth mentioning that there are more materials between the source and the detector when the source was placed at the bottom of the cryostat compared to that of the source at the top. Consequently, several low-energy X-rays and $\gamma$ rays ($\sim$20.8 keV and 26.3 keV) seen with the source at the top were not seen when the source was placed at the bottom. Instead, we observed two emitted X-rays with energy 24 keV and 27.5 keV, as shown in Fig.~\ref{fig:spec},  from indium due to the process of fluorescence when 59.5 keV X-ray interacted with the indium foil which is set underneath of the detector on the inside of the cryostat. To confirm the fluorescent emission of X-rays from the inner material of indium that is between the detector and the source from the bottom, we conducted a Geant4-based Monte Carlo simulation~\cite{geant4} for the source at the bottom with a detailed geometry of the cryostat based on our best understanding. Figure~\ref{fig:geant4} shows a comparison between the measurement and the simulation when the source is placed at the bottom. It was clear that two distinct X-rays observed in Fig.~\ref{fig:spec} when the source was placed at the bottom were from the fluorescent emission of indium. These two distinct X-rays provide a good opportunity to exam the detector energy resolution for low-energy X-rays with energy below 59.5 keV. 

\begin{figure} [htbp]
  \centering
  \includegraphics[clip,width=9.2 cm]{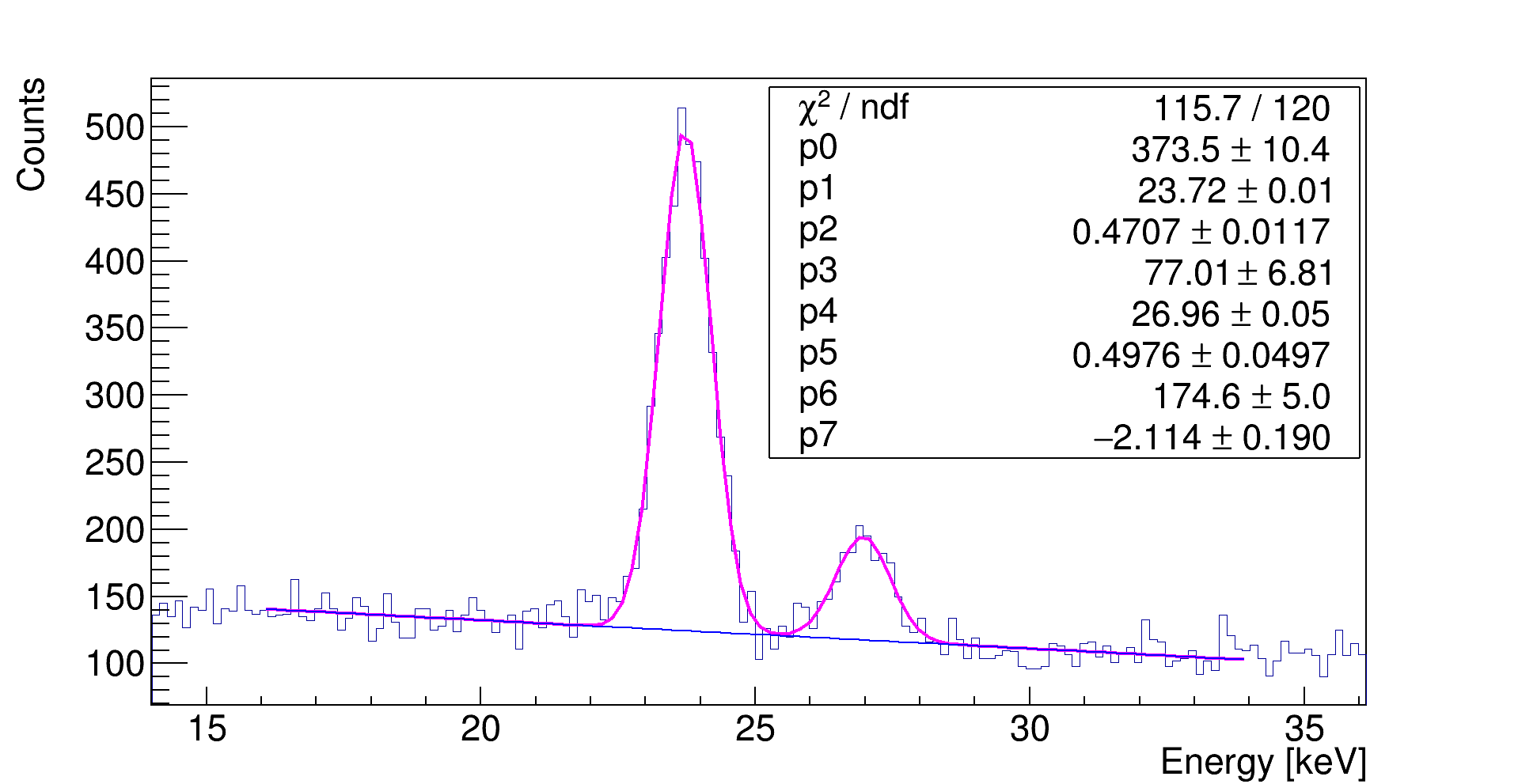}
  \caption{The fitted energy peaks at 24 keV and $\sim$27.5 keV from fluorescent emission of X-ray with applied voltage at 400 V when the source was placed at the bottom of the cryostat.}
  \label{fig:fit24_27}
\end{figure}

\begin{figure} [htbp]
  \centering
 \includegraphics[clip,width=9.2 cm]{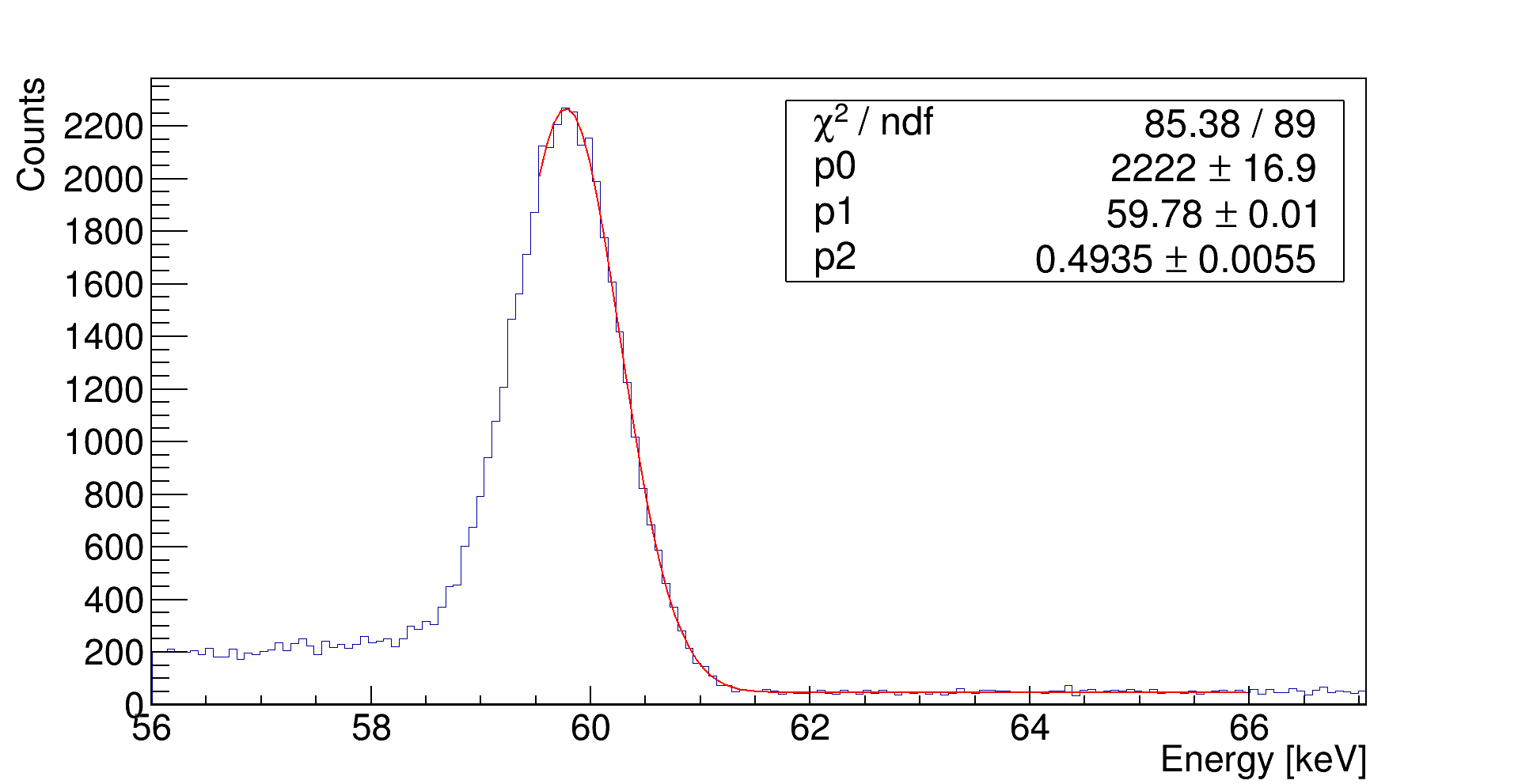}
  \caption{The fitted energy peak at 59.5 keV from Am-241 with applied voltage at 400 V when the source was placed at the bottom of the cryostat.}
  \label{fig:fit60}
\end{figure}

\begin{figure} [htbp]
  \centering
  \includegraphics[clip,width=9.2 cm]{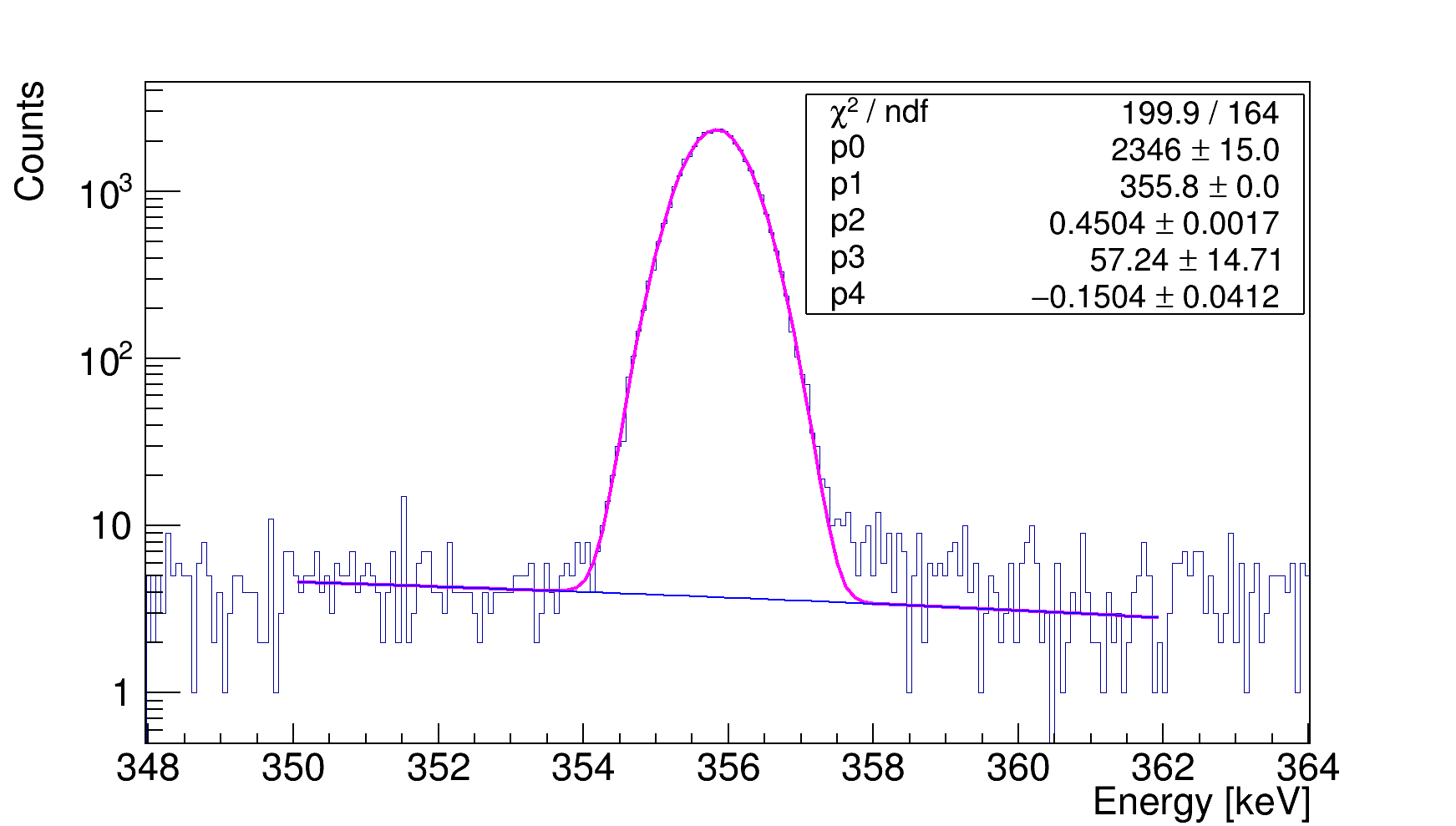}
  \caption{The fitted artificial peak due to the injected pulses from the high voltage line with applied voltage at 400 V when the source was placed at the bottom of the cryostat.}
  \label{fig:fitP}
\end{figure}

To obtain the detector energy resolution, the two low energy peaks at 24 keV and 27.5 keV were fitted together using two standard Gaussian distributions with a linear background distribution, $p0\cdot e^{[-\frac{1}{2}(\frac{E-p1}{p2})^2]}+p3\cdot e^{[-\frac{1}{2}(\frac{E-p4}{p5})^2]}+p6+p7\cdot E$, where $p0$ and $p3$ are the normalization constants, $p1$ and $p4$ are the center values in keV for the energy peaks at 24 keV and 27.5 keV, respectively, $p2$ and $p5$ are the Gaussian widths (sigma) in keV for the energy peaks at 24 keV and 27.5 keV, respectively, $p6$ is a constant, $p7$ is the slope, and $E$ is the energy in keV. The energy peak at 59.5 keV was partially fitted using a standard Gaussian distribution, $p0\cdot e^{[-\frac{1}{2}(\frac{E-p1}{p2})^2]}$, with $p0$, $p1$, $p2$ and $E$ defined above. This is because the background on the left side is higher than the right side and we are only interested in knowing the energy resolution. As examples, Figs.~\ref{fig:fit24_27} and ~\ref{fig:fit60} show the fitting of these three energy peaks at 400 V when the radiation source was placed at the bottom of the cryostat. Figure~\ref{fig:fitP} is a fitted artificial peak due to the injected pulses from the high voltage line in order to obtain the noise level. The electronic circuit presented in Fig.~\ref{fig:electronic} shows how the pulse was generated to measure the electronic noise of the detector. The peak in Fig.~\ref{fig:fitP} was fitted using a standard Gaussian distribution and linear background distribution, $p0\cdot e^{[-\frac{1}{2}(\frac{E-p1}{p2})^2]}+p3+p4\cdot E$, with $p0$, $p1$, $p2$ and $E$ defined above, $p3$ is a constant and $p4$ is the slope. Table~\ref{tab:res} summarizes the full width at half maximum (FWHM) of the three energy peaks and the pulser peak shown in Figs.~\ref{fig:fit24_27} $-$~\ref{fig:fitP}. Also shown in Table~\ref{tab:res} are the statistically driven energy resolutions for each peak. These are determined by subtracting in quadrature the FWHM of the injected pulse peak from the FWHM of the respective energy peaks.

\begin{table*}[t]
  \centering
  \caption{The FWHM of three energy peaks from Am-241 and the pulser peak at 400 V when the source was placed at the bottom of the cryostat. The energies presented here are the expected values. For each energy peak, the relative statistic driven energy resolution was determined by subtracting in quadrature the FWHM at the pulser from the FWHM at the energy peak and then divided by the energy.}
 \begin{tabular}{p{2.5cm}p{2.5cm}p{2.5cm}p{3cm}}
\hline
 Energy Peak (keV) & FWHM at the energy peak (keV) &FWHM at the pulser (keV) & Relative statistic driven energy resolution (\%) \\ \hline
 24.0 & 1.11 & 1.06 & 1.37     \\ \hline
 27.5 & 1.17 & 1.06 & 1.80   \\ \hline
 59.5 & 1.16 & 1.06 & 0.79\\ \hline
  \end{tabular}
  \label{tab:res}
\end{table*}

\section{Discussion and Expectation}
\label{sec:dis}
 According to the discussion in Section~\ref{sec:field}, if the internal charge amplification occurs in the spectroscopy measurements with Am-241, the energy peaks are expected to shift significantly as a function of applied voltage. Unlike a fraction of charge increasing due to the improvement of the charge collection efficiency as a function of electric field, the internal charge amplification multiplies charge production by more than a factor of 2 or higher depending on the applied electric field. This multiplication of charge production can significantly shift the peak positions. As shown in Fig.~\ref{fig:spec}, for both positions of the source, no evident energy position shift has been observed for each energy peak. This indicates that the charge internal amplification didn't take place in the detector with applied voltage at 400 V. 
 
\begin{figure} [htbp]
  \centering
  \includegraphics[width=1\linewidth]{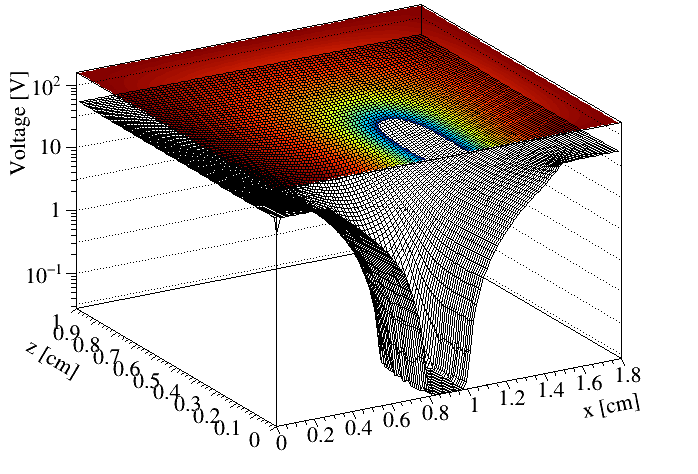}
  \caption{The potential field distribution in the detector when a positive voltage of 55 V was applied to the outside surface contact. The white area indicates that the detector is not fully depleted.}
  \label{fig:pot1}
\end{figure}

\begin{figure} [htbp]
  \centering
  \includegraphics[width=1\linewidth]{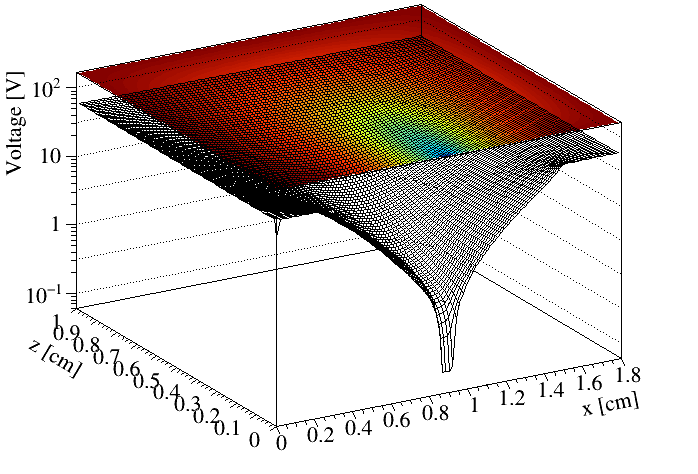}
  \caption{The potential field distribution in the detector when a positive voltage of 60 V was applied to the outside surface contact.The absence of any white area indicates that the detector is fully depleted.}
  \label{fig:pot2}
\end{figure}

\begin{figure} [htbp]
  \centering
  \includegraphics[width=1\linewidth]{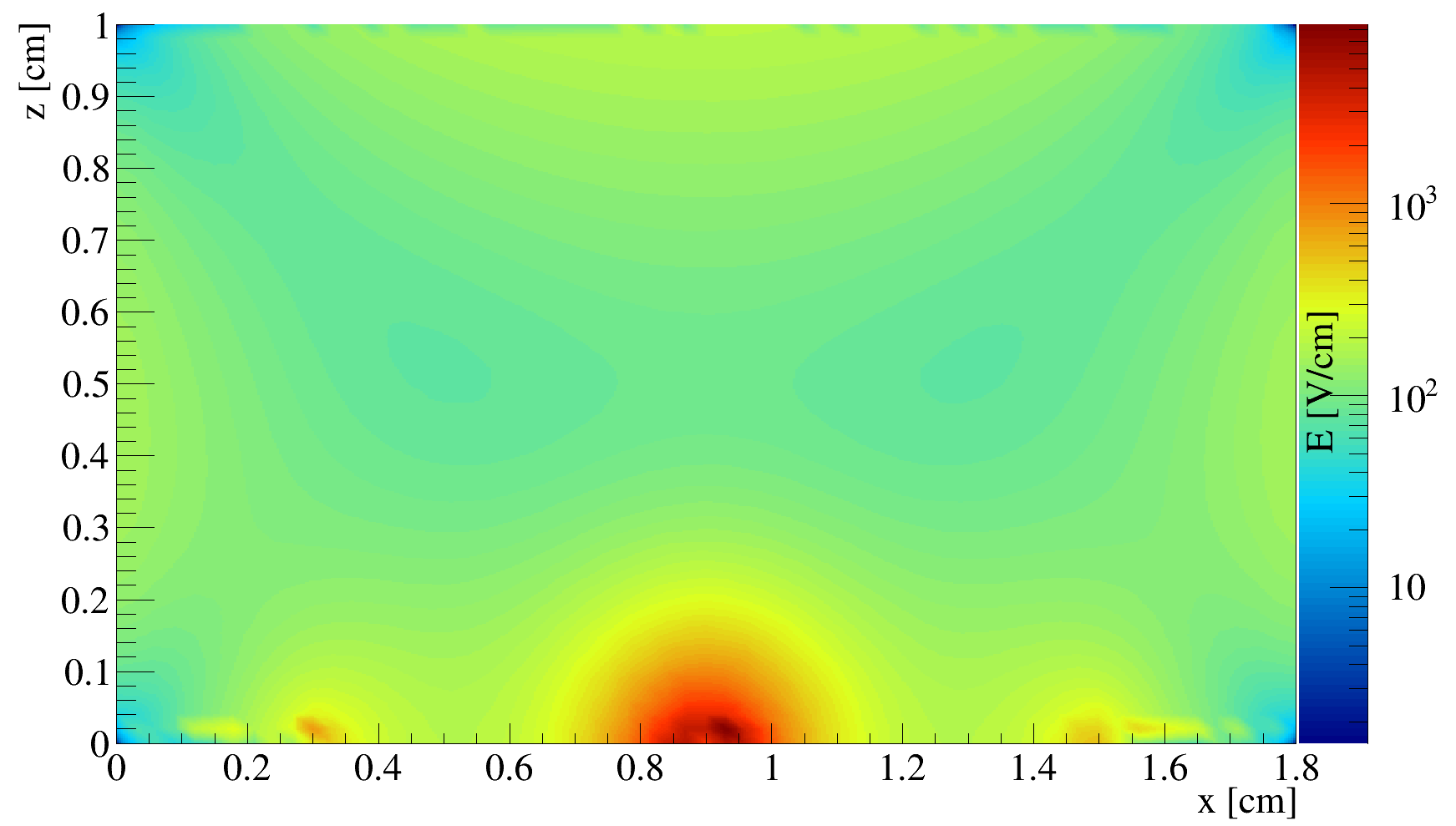}
  \caption{The electric field distribution in the detector when a positive voltage of 400 V was applied to the outside surface contact.}
  \label{fig:E}
\end{figure}

\begin{figure} [htbp]
  \centering
  \includegraphics[width=1\linewidth]{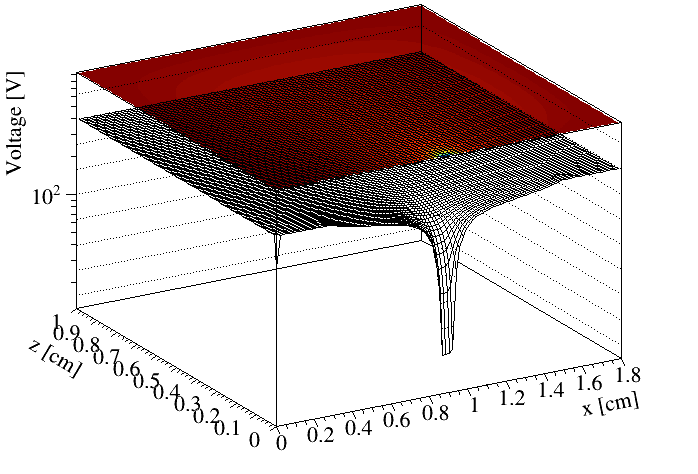}
  \caption{The potential field distribution in the detector when a positive voltage of 400 V was applied to the outside surface contact.}
  \label{fig:V}
\end{figure}

\begin{figure} [htbp]
  \centering
  \includegraphics[width=1\linewidth]{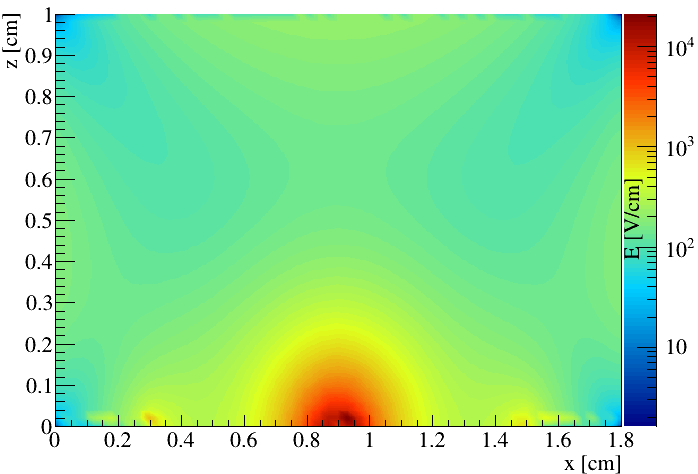}
  \caption{The electric field distribution in the detector when a positive voltage of 900 V was applied to the outside surface contact.}
  \label{fig:E900V}
\end{figure}

 From Section~\ref{sec:field}, the required electric field for the internal charge amplification to occur is at least $\sim$15,000 V/cm at 79 K. To find out the minimum applied voltage to allow our detector to achieve this required electric field, we have investigated the electric field distribution inside the detector using GeFiCa (Germanium Field Calculator)~\cite{Li}, which was created to demonstrate analytic and numeric methods to calculate static electric fields and potentials in HPGe detectors. A planar geometry with the same dimensions as the real detector (except four wings where the electric field is expected to be extremely low) has been built in GeFiCa. As shown in Figs.~\ref{fig:pot1} and ~\ref{fig:pot2}, the detector can be fully depleted at 60 V in GeFiCa and is in agreement with the measurement as shown in Fig.~\ref{fig:CV}. Given the inputs for the detector dimensions and depletion voltage (60 V) in GeFiCa, the impurity concentration of the crystal predicted by GeFiCa is 3.3$\times$10$^{9}$/cm$^{3}$, which is close to the measured impurity concentration of the two planar detectors (USD-W04 and USD-RL01)~\cite{mei2} made from the same crystal as the detector, USD-W06, in this work. This validates the predication of the impurity concentration by GeFiCa. Figures~\ref{fig:E} and ~\ref{fig:V} show the distributions of the electric field and potential inside the detector with an applied voltage of 400 V on the outside surface contact. As shown in Fig.~\ref{fig:E}, the maximum electric field in the detector at 400 V is between 8,000 V/cm and 9,000 V/cm, which is significantly lower than the required minimum electric field of $\sim$15,000 V/cm for internal charge amplification to occur inside the detector. This confirms that no internal charge amplification can be observed in our detector with an applied voltage of 400 V. Figure~\ref{fig:E900V} shows the expected electric field distribution inside the detector with an applied voltage of 900 V predicted by GeFiCa. This indicates that, to observe the internal charge amplification with our detector, the applied voltage needs to be at least around 900 V to reach the required electric field of $\sim$15,000 V/cm. Unfortunately, due to the fact that the leakage current of our detector jumped to a very high value starting from around 420 V because of field penetration into a-Ge, we were unable to apply the bias voltage beyond 400 V. For our future work, we will optimize our fabrication process to improve the quality of the a-Ge contact so that it will allow the detector to withstand the required electric field to observe the internal charge amplification. 
 
\section{Conclusions}
\label{sec:conl}
We have experimentally shown that a planar PPC detector can be fabricated successfully using USD-grown crystals with a-Ge contacts. The minimum required electric field to achieve internal charge amplification is derived using the work done on the charge carriers by the externally applied electric field. This work done must be equal to the sum of the kinetic energy gained by the charge carriers and the energy loss to the emission of photons. This relation is required by the energy conservation during the drifting process. From the calculation using GeFiCa, we find that the planar PPC Ge detector can be used to achieve the high electric field needed for triggering internal charge amplification. Such a detector, when operated at a bias voltage of higher than 900 V, will generate a field strength that is higher than 15,000 V/cm near the point contact. This high field will then amplify charge carriers to generate more charge carriers. We examined the energy resolution of the planar PPC detector and found that a good energy resolution can be achieved with the current electronics as shown in Table~\ref{tab:res}. If a low-noise electronic system is adapted and the energy resolution is dominated by the statistical variation, the energy resolution of 1.37\% at 24 keV, 1.80\% at 27.5 keV, and 0.79\% at 59.5 keV can be achieved. If internal charge amplification can be realized by this detector with an amplification factor of 1000, one can achieve a low energy threshold of $\sim$0.33 eV (1.37\%$\times$24 keV/1000). This suggests that planar PPC detectors can be used for low-mass dark matter searches. 

\begin{acknowledgement}
The authors would like to thank Mark Amman for his instructions on fabricating planar detectors, and the Nuclear Science Division at Lawrence Berkeley National Laboratory for providing the vacuum cryostat. We would also like to thank Christina Keller for a careful reading of this manuscript. This work was supported in part by NSF NSF OISE 1743790, DOE grant DE-FG02-10ER46709, DE-SC0004768, the Office of Research at the University of South Dakota and a research center supported by the State of South Dakota.
\end{acknowledgement}

\end{document}